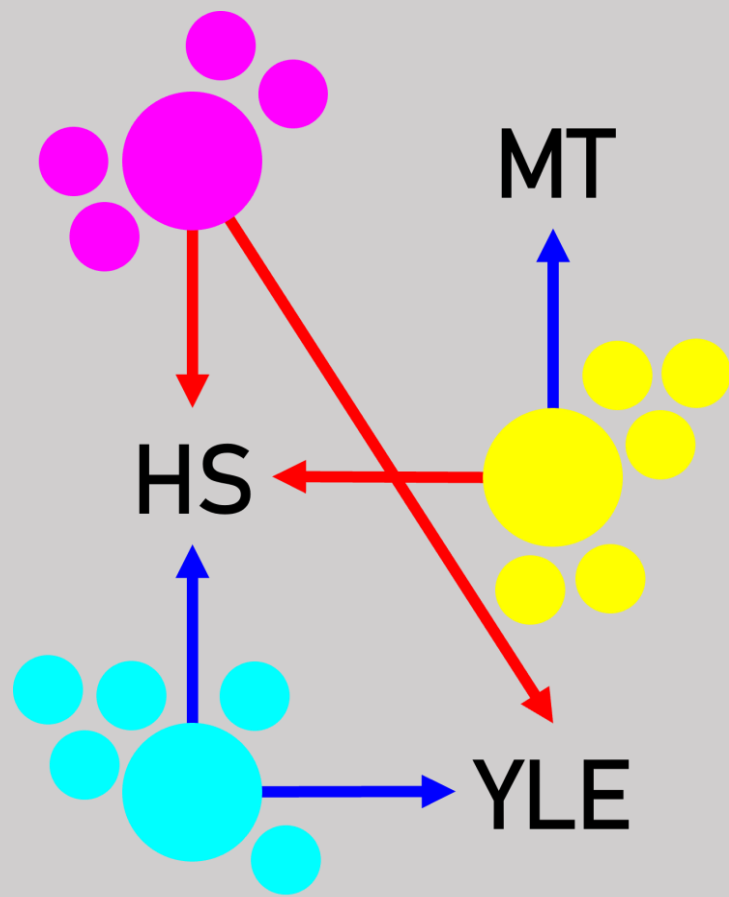

**Political polarisation in turbulent times**
Tracking polarisation trends and partisan news link sharing on Finnish Twitter, 2015–2023

**Technical report**
Helsingin Sanomat Foundation

**March**
2024

*Political polarisation in turbulent times*
Tracking polarisation trends and partisan news link sharing on Finnish Twitter, 2015–2023

*Technical report*
Helsingin Sanomat Foundation

*March*
2024

*Authors*
Antti Gronow*
Arttu Malkamäki*^

*Faculty of Social Sciences, University of Helsinki
^Department of Communication, Stanford University

*Contents*

*Summary*

The study analyses polarisation on Finnish social media with data from the platform *X*, which was known as Twitter during the time of data collection (during the Sipilä and Marin governments, 2015–2023). The users were clustered into three different ideological groups – the Conservative Right, the Moderate Right, and the Liberal Left – based on their retweeting of tweets referring to the different political parties in Finland. Trends in polarisation of several topics encompassing the most recent political crises – immigration, climate change, COVID-19, and security policy - between these ideological groups is analysed using network methods. To what extent the polarisation of each topic aligns with the polarisation of the other topics is also studied. In addition, the sharing of news links is examined in relation to the ideological groups of the users as well as to the sentiment and the virality of the tweets in which news links are shared.

    Our results show an upward trend in polarisation under most of the topics, especially during the Marin government. For all topics, the polarisation between the Conservative Right and the Liberal Left is high. The immigration topic is particularly dominated by active conservative-right users. The topic of COVID-19 differs from the other topics in that the Conservative Right and the Moderate Right are also highly polarised. In line with previous research, there is considerable alignment in the polarisation of the immigration and climate change topics. For COVID-19, alignment with the immigration and climate topics increases when the first COVID-19 vaccines were introduced. This alignment likely indicates that some users who are critical of immigration and climate action also start to retweet critical opinions of COVID-19 vaccines.

    The results for news link sharing show that all ideological groups rely on partisan media to a certain extent, which indicates that they selectively share news links that reinforce rather than challenge their pre-existing partisan opinions. The Conservative Right differs from the other groups in that it also shares links to fringe media outlets known for disinformation and far-right views. In addition, sharing links to the major news outlets in Finland, such as Helsingin Sanomat and YLE, is common among all groups. However, a large proportion of the Conservative Right share a different set of news articles than the other groups especially in relation to immigration. In addition, negative sentiments are over-represented in tweets where links are shared from within the Conservative Right. More generally, both positive and negative sentiments increase the virality of tweets in which links are shared but the Conservative Right is unique in that its members on average exhibit negative sentiment more often than the other groups. Furthermore, tweets in which negative sentiments are expressed in the context of link sharing get a boost in virality. The Conservative Right thus exhibits a "negativity bias" in the sense that tweets with negative sentiment gain considerable traction within this group.

*Tiivistelmä*

Tutkimuksessa analysoidaan suomalaisen sosiaalisen median polarisaatiota *X*-alustalla, joka tiedonkeruun aikana (Sipilän ja Marinin hallitusten aikana, 2015–2023) tunnettiin Twitterinä. Käyttäjät ryhmiteltiin kolmeen ideologiseen ryhmään – konservatiiviseen oikeistoon, maltilliseen oikeistoon ja liberaalivasemmistoon – sen perusteella, miten he uudelleentwiittasivat twiittejä, joissa viitattiin Suomalaisiin poliittisiin puolueisiin. Polarisaation kehitystä näiden ryhmien välillä analysoitiin useiden viimeaikaisten poliittisten kriisien - maahanmuutto, ilmastonmuutos, COVID-19 ja turvallisuuspolitiikka – ympärillä verkostomenetelmiä hyödyntäen. Tutkimme myös, missä määrin kunkin aiheen polarisaatio vastaa muiden aiheiden polarisaatiota. Lisäksi tarkastelemme uutislinkkien jakamista suhteessa niin käyttäjien ideologisiin ryhmiin kuin uutislinkin sisältävien twiittien tunnelataukseen ja viraalisuuteen.

Tulokset osoittavat, että polarisaatio lisääntyi lähes kaikkien aiheiden ympärillä Marinin hallituksen aikana. Erityisesti konservatiivioikeiston ja liberaalivasemmiston keskinäinen polarisaatio on korkealla. Maahanmuuttoaihetta puolestaan hallitsevat aktiiviset konservatiivioikeistolaiset käyttäjät. Aiheena COVID-19 eroaa muista aiheista siinä, että polarisaatio kasvaa myös konservatiivioikeiston ja maltillisen oikeiston välillä. Aiempien tutkimusten mukaisesti maahanmuutto- ja ilmastoaiheiden polarisaatiossa on havaittavissa huomattavaa samankaltaisuutta siten, että samat käyttäjät löytävät itsensä samoista, toisilleen vastakkaisista ryhmistä. Koronan osalta tällainen samankaltainen polarisaatio lisääntyy maahanmuutto- ja ilmastoaiheiden kanssa, kun ensimmäiset rokotteet otettiin käyttöön. Tulos viittaa siihen, että jotkut maahanmuuttoa ja ilmastotoimia kritisoivat käyttäjät ryhtyvät uudelleentwiittaamaan myös kriittisiä mielipiteitä koronarokotteista.

Uutislinkkien jakamisen osalta tulokset osoittavat, että kaikki ideologiset ryhmät nojaavat tietyssä määrin "puolueellisiin" medialähteisiin eli esimerkiksi puoluelehtiin. Tämä viittaa siihen, että käyttäjät jakavat valikoivasti sellaisia uutislinkkejä, jotka tukevat heidän poliittisia mielipiteitään. Konservatiivioikeisto eroaa muista ryhmistä, sillä sen jäsenet jakavat linkkejä disinformaatiosta ja ääriokeistolaisista näkemyksistä tunnetuille sivustoille. Linkkien jakaminen maan suurimpiin uutismedioihin (mm. Helsingin Sanomat ja YLE) on kuitenkin yleistä kaikissa ryhmissä. Osa konservatiivioikeistoa jakaa kuitenkin eri uutisia kuin muut ryhmät erityisesti maahanmuuttoon liittyen. Lisäksi negatiiviset sentimentit ovat yliedustettuina konservatiivioikeiston uutislinkkejä jakavissa twiiteissä. Positiiviset ja negatiiviset tunnelataukset yleisesti lisäävät uutislinkkejä jakavien twiittien viraalisuutta. Uutislinkkejä jakavien tweettien viraalisuutta tarkasteltaessa konservatiivioikeisto erottuu joukosta siten, että negatiiviset tunnelataukset lisäävät tweettien viraalisuutta tässä ryhmässä. Konservatiivioikeistolla on siis "negatiivisuusvinouma" siinä mielessä, että negatiivisia tunteita sisältävillä twiiteillä on taipumus päätyä viraaleiksi.

*Acknowledgements*

We thank Jussi-Veikka Hynynen, Ali Salloum, Sonja Savolainen, Ville Saarinen, Mikko Kivelä, Tuomas Ylä-Anttila, Jeffrey T. Hancock, and Risto Kunelius for their useful inputs on our work. The project was funded by a grant from the Helsingin Sanomat Foundation (#20210021).

*Suggested citation*

Gronow, A., Malkamäki, A. (2024). Political polarisation in turbulent times: Tracking polarisation trends and partisan news link sharing on Finnish Twitter, 2015–2023. Technical report.

# INTRODUCTION
## What is political polarisation

Political life today increasingly looks like a cascade of crises. From the 2008 global financial crisis to the 2015 European migrant crisis, the climate emergency, the COVID-19 pandemic, and the 2022 Russian invasion of Ukraine, political crises expose weaknesses in the ability of political actors and institutions to address complex challenges, encourage political opportunism, and evoke fear and uncertainty among citizens. In so doing, crises potentially polarise societies by pushing political groups further apart from one another. As most crises are at least partly experienced indirectly through news media, understanding the role of professional journalism in either increasing or mitigating polarisation matters. Moreover, many worry that social media exacerbates polarisation amid crises either by contributing to the rise of "echo chambers" or amplifying the more radical voices in society that often deviate from those heard in mainstream news media (Cinelli et al., 2021; Hong et al., 2019).

Polarisation has arguably been increasing across democracies (Boxell et al., 2020; McCoy et al., 2018; Wagner, 2020). Political polarisation, however, has many faces. Traditionally, it has referred to increasing distances in views on single issues or in relation to a range of issues (Sartori, 1976). In the latter case, stances on multiple issues correlate with each other and polarisation therefore relates to different ideologies. Some authors argue that issue-based polarisation is not a problem as long as opinions on all salient issues are not aligned in this way (DellaPosta, 2020). Ideological polarisation is more problematic because it can lead to entrenched ingroup identities, which create animosity against the political outgroup and make compromises across the political spectrum difficult (Mason, 2015).

A combination of ingroup praise and outgroup animosity, which does not necessarily anymore correspond to ideological polarisation, is referred to as relational, or affective, polarisation (Iyengar et al., 2019). It is associated with numerous societally negative outcomes, including poor governance amid crises (Rodriguez et al., 2022). Most studies in the United States have witnessed an upward trend in affective polarisation along partisan lines, to which changes in political communication brought about by social media over the past decade have arguably contributed (Bail et al., 2018; Frimer et al., 2023; Levy, 2021). There are some signs that affective polarisation may be on the rise also in Finland, although understanding the phenomenon is more complicated in a multi-party than in a two-party system (Kekkonen and Ylä-Anttila, 2021).

## What do we study

In this report, we cover the evolution of political polarisation on Finnish social media in the context of the most pressing political crises of the last decade and pay attention to what extent the polarisation that occurs in the context of a certain crisis aligns with polarisation amid other crises (Task 1). In addition, we explore how journalistic news stories and their sharing on social media potentially contribute to the polarisation of the discussion (Task 2).

One possible explanation for the upward trend in political polarisation is the mediatisation of politics – political institutions and the society having become increasingly dependent on media (Strömbäck, 2008). Particularly in the highly polarised United States, media-party parallelism has been strengthening over the past decades. This means that partisan media present overly biased political framing and commentary of political news. For the media consumer, they enable the curation of news consumption based on one's existing political beliefs. For example, partisan cable channels tailor their political coverage to persuade either Democrats or Republicans (Muise et al., 2022). By choosing which channel to watch, one can selectively expose oneself to news content that reinforces rather than challenges one's political views.

The contemporary media landscape in Finland is different. Lacking strong media-party parallelism, the public broadcaster (YLE) and a handful of commercial outlets (e.g., Helsingin Sanomat and MTV3) dominate the news market across online and offline mediums, with an overwhelming majority of Finns also trusting these media as news sources (Lelkes, 2016; Newman et al., 2023). In addition to the differences in media systems, Finnish political institutions also differ from their American counterparts. It is possible that the traditionally relatively consensual political institutions of Finland "safeguard" the society from the most pernicious forms of polarisation (Hallin and Mancini, 2004). For example, the Finnish multi-party system, which necessitates to compromises in coalition governments, may alleviate polarisation. One could even argue that the consensual political institutions of Finland make it an unlikely case for steep political polarisation. Recent trends, however, indicate increasing polarisation among political elites, especially on immigration and climate change policies (Lönnqvist et al., 2020).

Social media may boost polarisation, but it also opens a window into studying the most polarised segments of society. Previous research has shown that especially immigration and climate change politics exhibit polarisation on Finnish social media and that the resulting clusters of social media users are largely the same across these two topics (Chen et al., 2021). The analysis has however been confined to a single point in time around the 2019 parliamentary elections and has therefore not been able to judge whether and when polarisation in relation to the two topics has increased (or perhaps decreased). Some studies in a non-Finnish context have tracked polarisation over a longer period, for example, in relation to climate change on social media, by showing that the polarisation of climate politics has been increasing (Falkenberg et al., 2022). However, studies that have adopted a long-term perspective have predominantly concentrated on the polarisation of single issues.

In our Task 1, we map the polarisation of topics corresponding to four different political crises, and their interconnections, over a period of eight years, spanning the centre-right Sipilä government (2015-2019) and the centre-left Marin government (2019-2023). By topic, we refer to, for example, all discussion on immigration on social media. A topic may comprise multiple issues – consider, for example, the issues of racism and workforce dynamics under the topic of immigration. Our topics include immigration and climate change, but we also study to what extent the topic of COVID-19 polarised users on Finnish social media. In the United States, partisan affiliation has been found to correlate with both the policy preferences concerning the pandemic and the excess death rates among citizens (Rodriguez et al., 2022; Wallace et al., 2023). Although Finnish COVID-19 response was more consensual, social media, such as Twitter, gives a voice to citizens who on average hold more extreme political views than the general population (Barberá and Rivero, 2015). Therefore, while the COVID-19 response in Finland might have been less polarised, Finnish social media users' response could have been more polarised. In the early stages of this research project (February 2022), Russia invaded Ukraine, prompting unprecedented national unity in relation to the idea of Finland joining the North Atlantic Treaty Organisation (NATO). To understand how polarisation adapted to such an existential external, we included security policy among our topics. Lastly, to illuminate the extent to which crises-related topics are polarised, we also included the topic of inequality – a classic line of division along the left-right axis – as a "counterfactual" polarisation trend.

In addition to tracking polarisation on social media within and across these topics, we analyse how polarisation is associated with users sharing news stories during the two different governments (Task 2). To the best of our knowledge, understanding the ways in which social media interact with legacy media in the polarisation process has received limited attention in existing literature.

As news dissemination is increasingly governed by social media, researchers worry that these high-choice environments generate less of a societal glue than the more centralised conduits of the past (Moeller et al., 2016). We do not focus on news consumption per se but on the sharing of news links on social media. Some studies have investigated the sharing of news sources on social media and have found that the sharing of low-quality sources is more common among politicians (especially among Republicans) than among citizens in the United States (Greene, 2024). Our focus is on a phenomenon which, as far as we know, has not been studied before: how social media users of different ideological leanings share and frame news stories on social media. In the American context, where partisan media are important, one would expect to see partisans belonging to a different ideological group sharing news from different partisan outlets. As the media system and political institutions of Finland are different, partisan media are less likely to contribute to the creation of such segregated news environments even on social media.

However, it is still possible that the most partisan users rely on partisan media when they decide to share news because partisan media is likely to give support to their pre-existing vies. In addition, those who hold a strong partisan attitude are most likely to feel the urge to comment on news publicly on social media. Therefore, the role of those users that rely on partisan media is big in the case of sharing news links even though partisan media itself is not a big player in the Finnish media ecosystem. Relying on partisan media resembles the logic of selective exposure. The concept refers to an individual actively seeking and consuming information that aligns with one's existing preferences (Barberá et al., 2015). In the case of sharing links on social media, this means that only such links are shared that fit the political belief system of the user. It is likely to lead to selective exposure on the part of the users that are exposed to the shared posts either as followers or due to the platform's feed algorithm. We call this logic *selective sharing*.

The other option is that the role of partisan media in news link sharing is small and partisan users rely on mainstream media outlets. However, it is possible that the partisans' interpretation of mainstream media differs depending on their ideological stance. Thus, different social media users may frame the same links in different ways. In this case, sharing behaviour may look like it is not polarised in the sense that the same sources are shared but the accompanying social media posts in fact differ in their interpretation and framing. This constitutes a polarised setting, which is likely explained by a tendency for motivated reasoning and biased assimilation –the same news stories are interpreted and framed in a different light because of the partisan affiliation of the user (Bisgaard, 2015; Munro et al., 2002). Research on motivated reasoning and biased assimilation argues that when people encounter information contradicting their political views, they are likely to subject it to greater scrutiny and criticism (Lord and Taylor, 2009). Consequently, those with different ideological viewpoints could interpret the exact same news story differently. We call this logic *motivated reasoning*.

*Structural perspective on polarisation*
Rather than conceptualising polarisation merely as distance in opinions on political issues, scholars have recently introduced ways of measuring how individuals feel about other political groups (Iyengar et al., 2012). The polarisation that such measures capture, however, does not account for the social influence processes that effectively activate and structure polarisation (Jost et al., 2022). Like many political phenomena, polarisation is an outcome of the interactions among individuals, be they private citizens or members of the political establishment.

A network approach and its corresponding network analysis methods, which explicitly capture the interdependent nature of social interactions, constitute a structural, group-centric way of measuring polarisation (Salloum et al., 2022). In this perspective, users are clustered into political groups based on similar patterns of connectivity; the number and nature of concrete interactions within and between two or more groups are studied. Measuring polarisation in this way combines aspects of polarisation and makes it possible to map polarisation over time and across political topics (Task 1). A network approach also lends itself to studying processes that relate to polarisation through indirect interactions, such as users posting news media articles on social media (Task 2).

Until recently, the main obstacle to adopting a network approach for the study of political polarisation was the availability of reliable and meaningful data on typically momentary social interactions. With the advent of social media, however, the availability of data has improved.

METHODOLOGY
*Twitter data*
As social media play an increasingly central role in political agenda setting, politicians, journalists, activists, and private citizens increasingly rely on social media for political communication. Despite varying levels of algorithmic mediation, social media provide users with an affordable way of creating and responding to political content and tailoring the message to maximise reach and reaction, all without journalistic interference (Frimer et al., 2023). Scholars have also turned to social media as sources of rich data for providing insights into public opinion and political behaviour. Longitudinal social media data are ideal for monitoring how polarisation reacts in response to political crises and evolves over time.

We collected data from the social media platform Twitter (*X* since July 2023) over an eight-year period from 04/2015 to 03/2023. We used six sets of keyword queries, four of which relate to topics associated with recent political crises: the European migrant crisis (*immigration*), the climate emergency (*climate*), the COVID-19 pandemic (*coronavirus*), and the Russian invasion of Ukraine (*security*). The fifth set captures data relating to *inequality*. In addition, a sixth set contains keywords relating to the political *parties* in Finland. However, this set was used only to divide users into political groups based on their partisan-ideological interactions (for clustering based on statistical evidence, see Zhang and Peixoto, 2020). To interpret the resulting clusters, we made use of data containing the Twitter handles of parliamentary candidates in the 2019 and 2023 parliamentary elections.

The data include both original tweets and retweets (Table 1). As a form of interaction, retweeting – forwarding an original tweet without adding any commentary – is usually taken as endorsement because users tend to retweet political content with which they agree with and posts by users with similar political leanings (Barberá et al., 2015; Metaxas et al., 2015). We address Task 1 by constructing user--user endorsement networks from retweets and analysing structural polarisation by tracking retweeting behaviour under each topic and in relation to different pairs of political groups. A maximally polarised setting would be one where users assigned to a certain political group retweet exclusively among themselves on a given topic.

*Table 1.* Data by period, topic, and type (,000). *Note:* we omit coronavirus during the Sipilä government due to limited discussion on the topic.

| | Sipilä government 2015-04 >> 2019-03 | | | Marin government 2019-04 >> 2023-03 | | |
|---|---|---|---|---|---|---|
| | Tweets | Tweets with a news link | Retweets | Tweets | Tweets with a news link | Retweets |
| Immigration | 107 | 37 | 157 | 70 | 33 | 290 |
| Climate | 208 | 51 | 425 | 345 | 91 | 1,311 |
| Coronavirus | -- | -- | -- | 845 | 218 | 2,174 |
| Security | 135 | 40 | 199 | 253 | 73 | 871 |
| Inequality | 103 | 24 | 156 | 99 | 26 | 341 |
| Parties | 452 | -- | 599 | 601 | -- | 2,010 |

For our Task 2, we used original tweets that contained a hyperlink to news media, including mainstream, partisan, and "counter" news media outlets, whether online, print, radio, or television. By analysing user→news networks in which users interact indirectly via the news stories (and news stories via the users), we can evaluate the extent to which users from different political groups share similar news articles from similar media outlets, and whether their political leaning predicts the sentiment of their reaction to the news. This analysis also allows us to rank news stories (or users) under a given topic by their "virality". High virality results not only from widespread dissemination, but also from significant impact on public discourse.

*Partisan affiliation inference*
The clustering of users into groups based on how they retweet content relating to the Finnish political parties is shown in Figure 1. By dividing the users into two groups during the right-leaning Sipilä and the left-leaning Marin governments, and by inspecting the positions of the parliamentary candidates by political party (with the conventional party colour), we see a clear division along the government-opposition line. Therefore, supporters of the parties either in the government or in the opposition exhibit similar retweeting behaviour among themselves. When the government changes, this affects retweeting behaviour and the way in which the network is polarised. This makes sense, given that one would generally expect polarisation to occur between representatives and supporters of the government and the opposition.

However, polarisation is not necessarily limited to such formal, institutional divisions. By dividing the users into three groups, we observe a more partisan-ideological division into groups that correspond with the known blocs in Finnish politics along the left-right and liberal-conservative axes (Isotalo et al., 2020). Therefore, we label these three groups as Conservative Right, Liberal Left, and Moderate Right. Interestingly, we also see many candidates of minor parties (shown in orange in Figure 1), many of which held a conspiracist or far-right stance, contesting the 2023 elections and appearing in the Conservative Right group during the Marin government. Considering both the formal government-opposition and partisan-ideological groups allows us to draw a nuanced picture of polarisation trends. Hereafter, we refer to the two sets of groups simply as institutional and ideological, respectively.

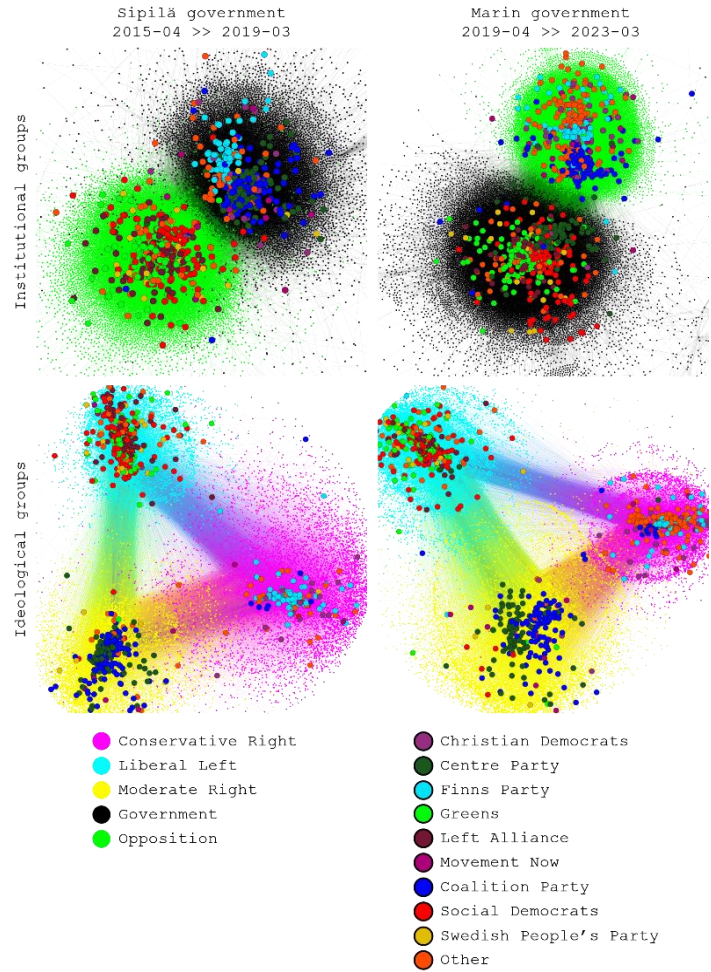

Figure 1. Institutional (Government and Opposition) and ideological groups (Conservative right, Liberal Left, and Moderate right) on Finnish Twitter during the Sipilä and Marin governments. Note: the black-framed nodes with varying colours refer to the candidates of major parties (orange collectively for minor parties) in the 2019 and 2023 parliamentary elections.

Tweet sentiment classifier

To classify the different framings accompanying the tweets, which include a link to a news article, we trained a tweet sentiment classifier with three classes (negative, neutral, and positive). Thus, we analyse the sentiment –negative, neutral, or positive expressions – of tweets, which contain a news link. The sentiment model made use of the FinnSentiment data containing 27,000 sentiment-annotated Finnish comments from Suomi24 (Lindén et al., 2023). The tweet classifier was based on a pretrained Finnish Megatron-BERT model (Kanerva et al., 2021). At 1.3 billion parameters, it is one of the largest natural language understanding models for the Finnish language. However, we had to fine-tune the model to accurately capture the sentiment in Finnish tweets, which appeared to be more negative on average than in the FinnSentiment data.

RESULTS

Task I: Trends in political polarisation in Finland

In Figure 2, we show how structural polarisation between the different political groups has evolved in relation to the five topics, and who participated in the discussions under each. The adaptive external-internal index (AEI) tracks polarisation in weekly intervals as the number of retweets (i.e., endorsements) within and between a pair of groups, while accounting for different group sizes (Salloum et al., 2022). The AEI score ranges between 1 and –1 with a higher positive number indicating higher polarisation. The lines in Figure 2 show polarisation between different pairs of groups and the bars at the bottom of each plot indicate how many members of each group participated in the discussion on a topic in each week. The share of users not classified as members of the groups (ideological or institutional) in Figure 1 over the eight-year period of our study is low, as indicated by the median count of such "non-partisan" users ranging between 12% and 18% across the five topics (immigration 12%, climate 18%, coronavirus 14%, security 13%, and inequality 12%). We take this as a validation of the representativeness of our partisan affiliation inference.

For the immigration topic, we see that a relatively low level of polarisation between the Conservative Right (ConRi) and Liberal Left (LibLe) (0.20-0.50) began climbing in 2018 (0.80-0.95) when many members of the Conservative Right joined the discussion (bars at the bottom of Figure 2 indicate the weekly count of active users by ideological group). At the time, the discussion concentrated on the controversial repatriation policies of the Finnish Immigration Service. Having already reached a high level before the 2019 parliamentary elections, both polarisation and participation levels have not reacted to the COVID-19 pandemic nor the 2022 Russian invasion of Ukraine. Perhaps surprisingly, very few Moderate Right (ModRi) members participate in the immigration discussion, and those who do, as shown by the low-lying LibLe-ModRi trend, often side with the Liberal Left. Consequently, the government-opposition (Gover-Oppos) trend closely resembles the ConRi-LibLe trend. The maximum user count within a single week is 873, which is less than for climate (2,888), coronavirus (7,589), or security (4,444). Thus, immigration is less popular than the other topics, which may partly relate to the fact that the keywords for this topic are more specifically defined than for the other topics (see Table A1 in the Appendix).

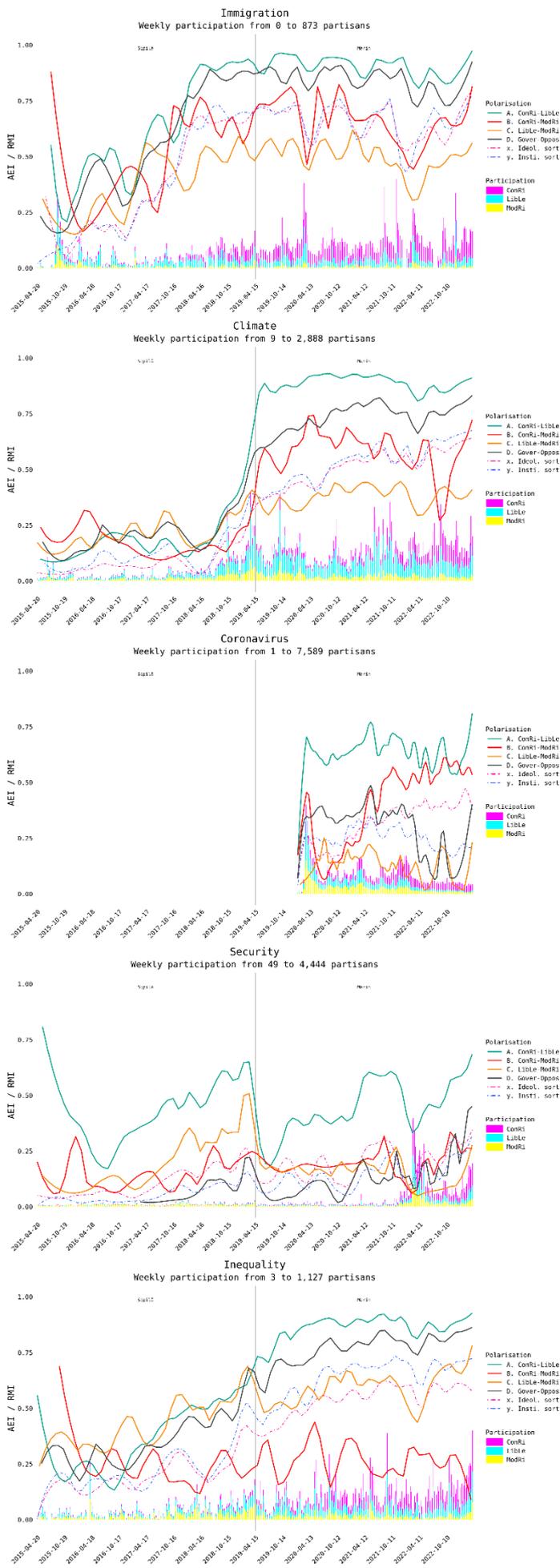

*Figure 2.* Polarisation, partisan sorting, and participation by topic in weekly intervals from 2015-04 to 2023-03

When we cluster users within each week either into one, two, or three groups, and again do so according to statistical evidence (Zhang and Peixoto, 2020), we can also track partisan sorting – the degree to which users stick to their assigned institutional and ideological groups under a given topic on a weekly basis – using the reduced mutual information index (RMI) (Newman et al., 2020). The RMI score ranges between 1 and 0 (possible negative values resulting from associations weaker or comparable to what would be expected at random fixed to 0). As the dashed lines in Figure 2 show, partisan sorting has been a decisive feature of the immigration discussion since 2018, as evident in the strong alignment of the weekly topic-specific partitions with the institutional (Instit. sort) and the ideological (Ideol. sort) groups (0.55-0.80). Both partisan sorting curves correlate strongly as most Moderate Right users remain silent on immigration – thus, members of the Conservative Right and the Liberal Left who do participate are largely reduced into representatives of the Government and supporters of the Opposition. This means that the political identity of a user would have been a good predictor of the user's retweeting behaviour under the immigration topic, but only if the user would have identified either with the Conservative Right or the Liberal Left.

The climate topic starts to polarise only in 2019 (following the publication of the so-called 1.5°C report by the Intergovernmental Panel on Climate Change), with the polarisation between the Conservative Right and the Liberal Left soon reaching a high level. The Conservative Right and the Moderate Right do not always see eye-to-eye when it comes to climate change but polarisation between these groups exhibits fluctuations. There was a drop in the polarisation of these two groups in relation to climate change around the middle of 2022, but the upcoming parliamentary elections clearly reverse this trend at the end of 2022. The Liberal Left has been very active in the discussion but the Conservative Right also becomes very active around 2022. Again, users' institutional (government versus opposition) and ideological background are closely correlated and increasingly polarise starting from 2019.

The discussion around COVID-19 was naturally non-existent before the end of 2019, as the disease was first identified in December 2019. Retweeting around the topic is very active in 2020 but there are also later peaks of activity. All three ideological groups are active at first but interestingly the Conservative Right is the only group which stays relatively active when general activity around the topic starts to fade in 2022. This finding may result from the interest of conspiracy theorists in the topic, as such actors are found in the Conservative Right group. The polarisation between the Conservative Right and the Liberal Left is high throughout the period but polarisation of the Conservative Right and Moderate Right increases in a linear fashion. This indicates that the Conservative Right is increasingly in disagreement with the Moderate Right. Further evidence comes from the finding that the even though there is some fluctuation in the polarisation between the Moderate Right and the Liberal Left, the polarisation of these groups stays on a relatively low level. Furthermore, unlike with the topics of immigration and climate, the lines for institutional and ideological sorting diverge as the pandemic progresses. This means that polarisation increasingly exists between ideological groups, not between the government and the opposition as such. Overall, the behaviour of the Conservative Right is unique when it comes to COVID-19.

Discussion around the security topic is almost non-existent before the end of 2021, which is when the possibility of Russia invading Ukraine became a salient issue. After the invasion in February 2022, all groups actively engage in retweeting behaviour. Polarisation between the Conservative Right and the Liberal Left stands out by being on a relatively high level. However, all polarisation curves start to rise as the Marin government is nearing its end and elections are approaching.

As for the topic of inequality, activity starts to increase nearing the end of Sipilä government. It is interesting that the Conservative Right becomes active in relation to this topic during Marin's government. The data reflects the fact that both the Conservative and Moderate Right represent right-wing views; the polarisation between these two groups hangs consistently at a lower level than the polarisation between the other pairs of groups. In addition, as the parliamentary elections of 2023 approach, polarisation between the Conservative Right and the Moderate Right drops. One could argue that the representatives and supporters of the right-wing populists and traditional right-wing interests find each other before the elections in 2023. The divergence of the ideological and institutional lines before the elections also points in this direction in that the Conservative Right and the Moderate Right come closer to one another. This is fitting in the sense that right-wing populists, namely Finns party, and the moderate-right wing party, Coalition party, joined hands in the new government in 2023.

Next, we shift our focus from the polarisation of single topics to the alignment of the weekly partitions within and across topics. We analyse the extent to which the clustering into groups in relation to each topic aligns with the clustering of the other topics during every week. In Figure 3, we see how the overlap tends to be larger within than between topics, and it generally increases towards the end of the Marin government. This makes sense as it shows that users participating in a topic rarely change from one group to another over time. As previous research has shown (Chen et al 2021), there is also considerable overlap between the immigration and climate themes, which dates to 2018. For the coronavirus topic, there is a likely change in the group structure, as alignment with the immigration and climate topics increases. This alignment coincides with the introduction of the first COVID-19 vaccines (the end of 2020 and beginning of 2021).

This alignment may indicate that some those who were presenting critical views of immigration and climate mitigation also started retweeting critical opinions of COVID-19 vaccines.

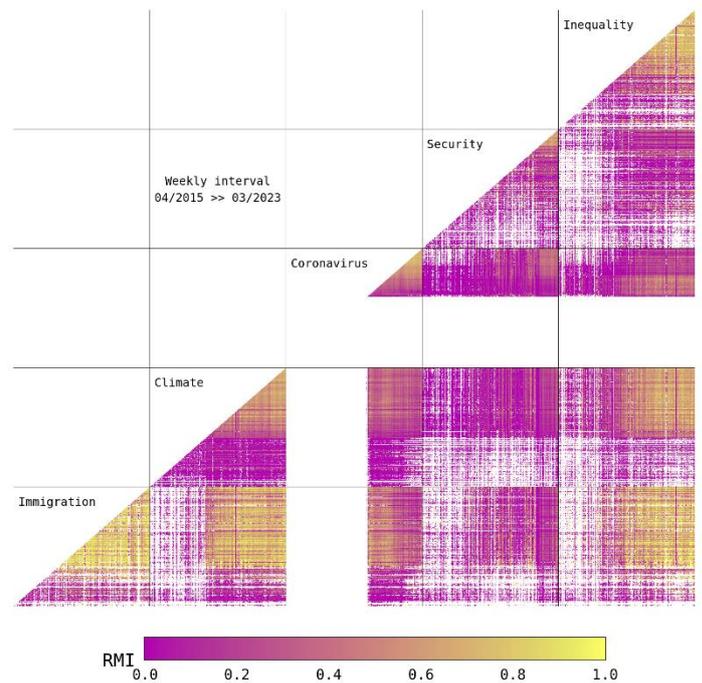

Figure 3. Alignment of weekly clustered groups within and across themes from 2015-04 to 2023-03. Note: cells without colour indicate that there was not enough statistical evidence for the existence of more than one group in at least one of the two weeks.

Interestingly, the overlap between the coronavirus and security themes stays surprisingly high (0.50-0.65) after the 2022 Russian invasion of Ukraine. This overlap may result from conspiracy theories gaining traction, with individuals who were sceptical of official narratives regarding the pandemic (including vaccines) also becoming suspicious of information surrounding the war in Ukraine. We also note that the inequality topic aligned with the immigration and climate topics already in 2018 (0.50-0.80). While the security topic exhibits only weak alignment with the other three topics, the COVID-19 topic aligns with immigration in 2020. This potentially stems from political opportunism during the first year of the pandemic when the discussion focused on the closing of borders – traditionally demanded by those who vocally oppose immigration.

Task II: How partisans share news links

We begin examining how the sharing of news media links is related with partisan differences by taking a closer look at the media outlets whose news stories are most often included in tweets circulating within each ideological group. This serves to establish whether users selectively share news from different outlets according to their partisan leanings.

The frequencies in Table 2 show that news links to the main news outlets, namely the public broadcaster YLE and Helsingin Sanomat, as well as the main tabloids Ilta-Sanomat and Iltalehti, are shared within all groups, across all themes, and over both periods (Sipilä and Marin governments). However, we also see that partisan media outlets receive considerable attention within the ideological groups. For example, Demokraatti, Vihreä Lanka, and Kansan Uutiset – all party-affiliated papers – circulate among the Liberal Left group where the Social democrats, the Greens, and the Left alliance are well represented. Links to Verkkouutiset, the voice of the Coalition party, are mostly shared by the Moderate Right, although Verkkouutiset has recently been shared also by the Conservative Right, especially in relation to the security and inequality topics. Suomen Uutiset, a Finns party news media, is well represented in the Conservative Right – but only for the immigration and climate topics, both of which are core policy areas of the party in relation to which the party often disagrees with the other main parties. Interestingly, the most popular outlet among the Conservative Right during both Sipilä and Marin governments is a fringe ethno-nationalistic online media Kansalainen. In addition, TokenTube, an online platform which brought various fringe communities in Finland together during the pandemic, ranks as the fifth most popular news outlet in the Conservative Right for the coronavirus theme.

The analysis above suggests that users mostly turn to the same main sources of information instead of exclusively relying partisan outlets. In other words, most users follow what the main media outlets are saying and do not exclusively live in an ideological "echo chamber." However, partisan sources also appear in Table 2. In addition, the Conservative Right stands out in that it relies on fringe sources of information for immigration, climate and COVID news, thus exhibiting a clear partisan media effect. Even if users do not selectively share news media from partisan news sources, the possibility of motivated reasoning still exists in that the linked news articles may be interpreted systematically differently by different groups. To find out whether such motivated reasoning is prevalent, we analyse the extent to which the sentiments in the tweets, where news media are shared, are determined by the ideological groups. Here, we report findings concerning tweets where the sentiment likely relates to the subject matter of the news link, not to journalism as such, as we filtered out tweets with keywords likely to be related with journalism (for tweets where the sentiment relates to journalistic features, see Figure A4 in the Appendix).

Table 2. Count of unique news articles retweeted by topic and ideological group. *Note:* partisan media outlets in italic.

| Sipilä 2015–2019 | Conservative Right | Liberal Left | Moderate Right |
|---|---|---|---|
| Immigration | *kansalainen.fi 167*<br>mtv.fi 148<br>yle.fi 148<br>il.fi 89<br>hs.fi 85 | yle.fi 144<br>hs.fi 84<br>il.fi 24<br>mtv.fi 22<br>suom.kuvalehti.fi 14 | yle.fi 50<br>*verkkouutiset.fi 39*<br>hs.fi 36<br>il.fi 21<br>kl.fi 15 |
| Climate | yle.fi 193<br>hs.fi 69<br>mtv.fi 48<br>is.fi 25<br>il.fi 18 | yle.fi 389<br>hs.fi 268<br>tekn.talous.fi 112<br>*vihrealanka.fi 94*<br>kl.fi 86 | yle.fi 226<br>hs.fi 156<br>talouselama.fi 107<br>kl.fi 104<br>mt.fi 65 |
| Security | mtv.fi 139<br>yle.fi 124<br>is.fi 101<br>uusisuomi.fi 60<br>hs.fi 58 | hs.fi 20<br>yle.fi 20<br>is.fi 9<br>*ku.fi 8*<br>il.fi 7 | *verkkouutiset.fi 936*<br>yle.fi 835<br>hs.fi 744<br>il.fi 354<br>is.fi 272 |
| Inequality | yle.fi 52<br>hs.fi 29<br>mtv.fi 29<br>uusisuomi.fi 23<br>il.fi 11 | yle.fi 125<br>hs.fi 67<br>*ku.fi 51*<br>is.fi 30<br>il.fi 22 | kl.fi 102<br>yle.fi 73<br>hs.fi 67<br>*verkkouutiset.fi 60*<br>talouselama.fi 51 |
| Marin 2019–2023 | Conservative Right | Liberal Left | Moderate Right |
| Immigration | *kansalainen.fi 239*<br>il.fi 90<br>is.fi 60<br>yle.fi 39<br>*suomenuutiset.fi 36* | hs.fi 38<br>yle.fi 25<br>*demokraatti.fi 4*<br>il.fi 3<br>mtv.fi 2 | mt.fi 2<br>hs.fi 1<br>*verkkouutiset.fi 1*<br>--<br>-- |
| Climate | is.fi 221<br>il.fi 185<br>mtv.fi 85<br>yle.fi 59<br>*suomenuutiset.fi 56* | hs.fi 458<br>yle.fi 412<br>mt.fi 83<br>il.fi 42<br>ts.fi 40 | yle.fi 38<br>talouselama.fi 15<br>hs.fi 12<br>mt.fi 12<br>kl.fi 10 |
| Coronavirus | is.fi 1,089<br>il.fi 737<br>mtv.fi 374<br>yle.fi 135<br>*tokentube.net 67* | hs.fi 607<br>yle.fi 598<br>*demokraatti.fi 204*<br>is.fi 69<br>il.fi 65 | is.fi 465<br>yle.fi 444<br>il.fi 245<br>hs.fi 240<br>mtv.fi 165 |
| Security | is.fi 600<br>il.fi 400<br>mtv.fi 227<br>*verkkouutiset.fi 94*<br>hs.fi 67 | hs.fi 303<br>yle.fi 192<br>*demokraatti.fi 112*<br>il.fi 22<br>is.fi 22 | hs.fi 89<br>*verkkouutiset.fi 84*<br>yle.fi 84<br>is.fi 76<br>il.fi 74 |
| Inequality | is.fi 37<br>il.fi 28<br>*verkkouutiset.fi 14*<br>yle.fi 10<br>hs.fi 7 | hs.fi 44<br>yle.fi 26<br>ku.fi 12<br>*demokraatti.fi 8*<br>il.fi 3 | kl.fi 7<br>talouselama.fi 7<br>*verkkouutiset.fi 5*<br>is.fi 4<br>il.fi 2 |

A vast majority of the tweets containing a news media link do not include a distinguishable negative or positive sentiment. The share of tweets with neutral sentiments ranges from 80% for immigration to 92% for climate, both during the Sipilä government. However, it must be considered that our tweet sentiment classifier was tuned to be conservative – that is, to avoid labelling unclear or ambiguous expressions as positive or negative. Nevertheless, those tweets that contain either positive or negative sentiments reveal interesting differences between the ideological groups. Figure 4 shows the structure of the user-news network emerging from link tweeting behaviour for the immigration, security, and COVID topics during the Marin government (the figures look similar during the Sipilä government, which is why we are not reporting these results here). The idea is that users interact indirectly by sharing news links and by exhibiting a positive or negative sentiment in the accompanying tweet (neutral sentiments are hidden from the figures for the sake of clarity but they affect the structure of the network). Thus, the sentiment relates to the way in which the users frame the shared link. In Figure 4, red links refer to negative and blue links to positive sentiments. To enhance the visibility of the sentiment on the left side of the figure, the nodes of the bipartite networks (i.e., users connect via news links, news links via users) are detached from (i.e., plotted above) the connections.

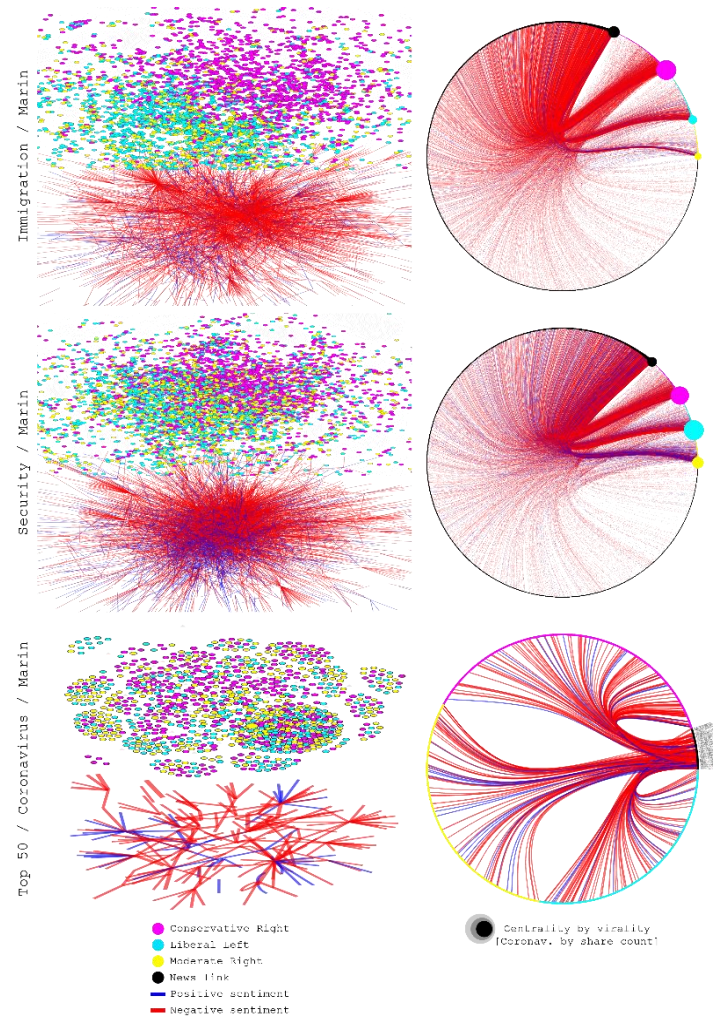

*Figure 4.* Sharing links to news articles on immigration during the Marin government. *Note:* red and blue lines refer to negative and positive sentiments in a tweet that comments the subject matter of a news story; neutral sentiments hidden.

On the left side of Figure 4, we see how members of the Liberal Left, the Moderate Right, and a fair share of the Conservative Right tweet news links in relation to the immigration, security, and COVID-19 topics. A large proportion of the Conservative Right, however, shares a completely different set of news outlets and articles than the other groups. In a bipartite network structure, this is evident due to the clearly visible clustering of Conservative Right, especially for the immigration topic, as Figure 4 shows. If selective sharing would *not* occur, partisans of all ideological leanings would cluster evenly around the same news links. In addition, negative sentiments (shown in red) cluster closely together, which means that sentiments are related with ideological groups. For the security topic, such clustering occurs to a lesser extent, although a similar division into the two camps persists. Despite the major external threat that led Finland to apply for NATO membership after a rapid surge in national unity following the 2022 Russian invasion of Ukraine (Xia et al. 2024), cohesion around security affairs in general appears weaker when it comes to sharing security-related news on Twitter. The topic of COVID-19, for which we here consider only a subset of the 50 most viral news links (see below), is messier than the other two topics and no clear clusters appear on the left side of the figure. This suggests that news dealing with COVID-19 generally need to attract attention across different ideological groups to gain significant traction on Finnish Twitter.

On the right side of Figure 4, we display the same information as on the left side in a format that clearly shows differences in sentiments and sorts both users and news links along a circle based on their "centrality" in the user-news network. Here, we operationalise centrality via the sum of incoming links to news stories (->news) and outgoing links based on the news that users share (user->), which have been weighted by virality. Virality is estimated based on the recommendation algorithm that Twitter (2023) recently made public (virality = 30 * like count + 20 * retweet count + 1 * reply count). By considering virality in the context of news link sharing, we can consider the extent to which other users react to the tweets where links are shared with accompanying sentiments. In Figure 4, the colour of a connection refers to the sentiment (red indicating negative and blue positive sentiments; neutral sentiments are hidden), width of connection to the virality of the tweet, and the size of the node to the centrality of the news story (black nodes), with users related with the three ideological groups as cyan, magenta, and yellow nodes.

Two main conclusions can be drawn from the right side of Figure 4. For the immigration topic during the Marin government, members of the Conservative Right share more links than the other two groups. However, most of their tweets do not become very viral. As indicated by the relative lack of sentiment in the connections leading to news stories with low aggregate virality, positive and negative sentiments usually increase virality. Sharing links to immigration news by displaying a negative sentiment is very common for the Conservative Right, while both the Liberal Left and Moderate Right exhibit more varied sentiments in this context. However, the Liberal Left also tends to favour negative more than positive sentiments when it comes to immigration. Potential reasons for the prevalence of negative sentiments among the Conservative Right include the notion of conservative-right voters being often critical of anything immigration-related. The social reward structure of this group may therefore play a role in that negative tweets may gain more reactions (Ceylan et al., 2023), and users may also adapt their behaviour based on feedback to maximise virality for their cause (Frimer et al., 2023). However, their behaviour is also a likely indication of motivated reasoning in that they tend to frame their news commentary in relation to immigration in a much more negative light than other users. We suspect that the negative sentiments displayed by the Liberal Left in this context do not indicate that they would think of immigrants in a negative manner. It is more likely that they share news where the ill treatment of immigrants is reported and therefore display negative sentiments.

From the relative sizes of the user nodes in Figure 4, we can see that relatively few actors in each ideological group gains attention for their tweets. In other words, few prominent users are responsible for tweeting and commenting news links that end up becoming viral. This result follows the logic of preferential attachment – that is, the initially higher popularity of a user leads to even higher popularity in the future, especially in online networks (Kunegis et al., 2013).

For the security topic, we see a similar pattern in terms of relatively few articles gaining virality, relatively few users being prominent in terms of the aggregate virality of their tweets, and the Conservative Right negatively framing news stories more often than the Moderate Right and the Liberal Left. In addition, tweets from the Liberal Left are frequently accompanied by a positive sentiment. This contrast is surprising in the sense that one would expect the Liberal Left to be more critical of security issues than right-wing actors. In addition, security affairs are the traditional focus area of the moderate-right Coalition party, but on social media the Conservative Right is still more likely to share news links on this topic on Twitter. Some members of the Moderate Right nevertheless attract considerable attention with their tweets.

By focusing on the top tweets with news links for COVID-19, we may affirm that a news story usually becomes viral when it attracts attention across ideological lines, even though there are a few notable exceptions. The news article that tops the list is a Helsingin Sanomat feature article focusing on Mika Salminen – former head of health safety at the Finnish Institute for Health and Welfare (THL) – who criticises the national response to the COVID-19 pandemic (Figure 5). The link has been tweeted by 16 (pos: 2, neg: 1) conservative-right, 16 (pos: 3, neg: 2) liberal-left, and 33 (pos: 2, neg: 4) moderate-right users. Beyond the mere count of tweets, the virality of the tweet (rounded to thousand) is 78 (pos: 0, neg: 55) for the Conservative Right, 27 (pos: 0, neg: 0) for the Liberal Left, and 48 (pos: 0, neg: 4) for the Moderate Right groups. The distribution of sentiments within each group provides grounds for suspecting that motivated reasoning governs the way in which partisans interpret the same piece of news. Notably, the positive sentiments among liberal-left users do not evoke any virality, while even a single negative sentiment by a conservative-right partisan becomes viral. Thus, even though the Liberal Left is positive in many of their sentiments in relation to security issues, the response to such tweets in this group is relatively muted. In addition, the Conservative Right exhibits a "negativity bias" in the sense that the negative sentiments expressed in this group in relation to security issues tend to become viral.

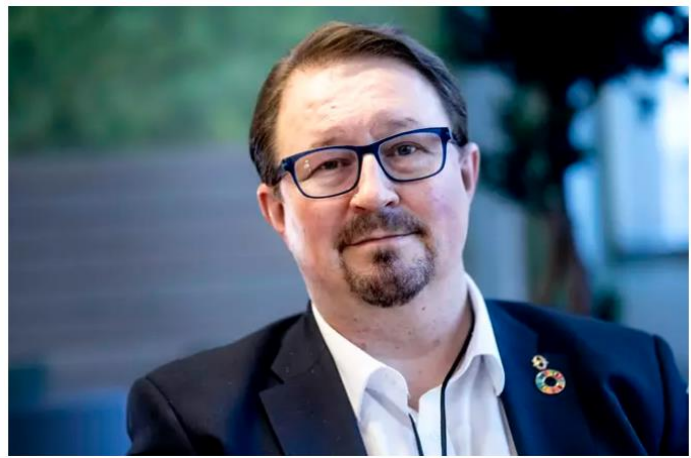

Figure 5. The most viral news story on Finnish Twitter under the COVID-19 topic during the Marin government.

The second most viral article is a "live coverage" article by YLE that reports the introduction of the so-called COVID-19 vaccination passport in the European Union and the extension of the prohibition of in-class teaching in Finnish universities. It has been tweeted by 77 (pos: 2, neg: 9) conservative-right, 85 (pos: 2, neg: 9) liberal-left, and 91 (pos: 3, neg: 10) moderate-right users, with virality at 15 (pos: 0, neg: 4), 18 (pos: 0, neg: 1), and 115 (pos: 0, neg: 39). The news link has collected shares from users belonging to different ideological groups because the story presumably has something for everyone; the Conservative Right can pay attention to the issue of the vaccination passport (likely by disagreeing with the idea) and the Liberal Left to the criticism of the decision to prolong distance education because of its negative impact on students. Once again, we observe how even a relatively few tweets with a negative framing of a news article can attract attention on social media and potentially fuel ingroup emotional convergence within the ideological groups (Parkinson, 2020).

While almost all of the 50 most viral news under the coronavirus topic originate from mainstream legacy media, there are two articles from Seiska, an outlet representing sensationalistic "yellow" press, which report on the actions (and the consequences of those actions) of the foreign secretary Pekka Haavisto (Figure 6) and Sanna Marin, prime minister at the time, which led to their exposure to COVID-19. Attention to both articles concentrates among the Conservative Right. Out of the 30 Conservative Right tweets sharing this article, 11 are classified as negative. More interestingly, these 11 negative reactions contribute 82% of the total virality (120) of the article. Thus, sharing news links with negative framings may not always be the most common way of sharing news links but negativity pays off especially in the Conservative Right.

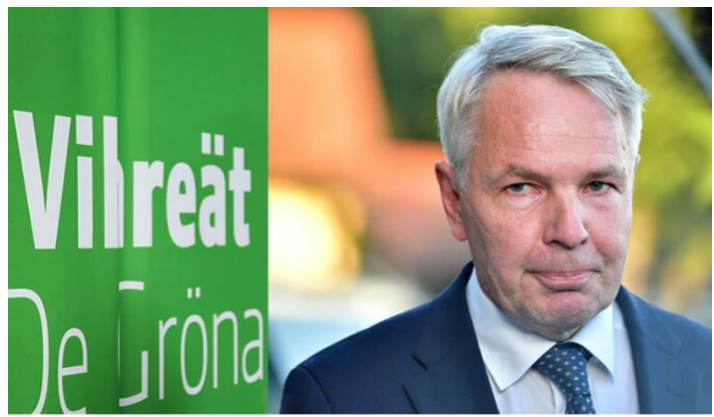

*Figure 6.* The fourth most viral news story on Finnish Twitter under the coronavirus topic during the Marin government.

Here, we have chosen to report some of the most interesting examples of our descriptive results. However, the corresponding results for the other topics during the two different governments, as well as for the tweets targeting the journalistic features of the news story, largely exhibit similar behavioural patterns.

CONCLUSION
When analysing polarisation from a network perspective, the distinction between an ingroup and an outgroup is essential. This is because polarisation is assessed by analysing the way in which interactions between different groups are distributed. As we have seen, the more groups only interact among themselves, the more polarised the setting. With social media data, attribute data on the users is usually lacking in the sense that it is difficult, for example, to ask the users which party they voted for. Previous network-based polarisation analyses done on social media data have, for example, used moving time windows where changes in retweeting behaviour in relation to the topic in question are used to assess polarisation levels (Chen et al., 2021; Salloum et al., 2022; Xia et al., 2024). While doing so is fine in terms of capturing changes in retweeting behaviour, it is often more interesting to see how users with different ideological backgrounds interact in relation to different kinds of topics. We introduced a method for clustering users into ideological groups based on their retweeting of tweets citing one or more of the main political parties in Finland (separately for 2015-2019 and 2019-2023) and used this clustering as a basis from which to infer polarisation trends for four different topics, all related with the main crises of our times.

Our results gave clear indications that polarisation on Finnish Twitter in relation to the studied topics has increased in recent years, and especially during the Marin government. The polarisation between the Conservative Right and the Liberal Left stands out by being at high level across all topics. This result is not surprising in the sense that the representatives and supporters of right-wing populists and the left and the greens have often been in opposition with each other in Finnish politics in recent years. However, the Conservative Right stands out also in other ways in that the Conservative Right and the Moderate Right also highly polarised during the COVID-19 pandemic. In addition, the polarisation of traditional topics of concern, namely immigration and climate change issues, starts to align with the coronavirus topic around the time when the first COVID-19 vaccines were introduced. This alignment likely indicates that some conservative-right users critical of immigration and climate mitigation also start to retweet critical opinions of the vaccines. Thus, even if the right-wing populist Finns Party did not officially start to question COVID-19 vaccines – unlike President Trump in the United States – some social media users that share populists' concerns with immigration and climate change mitigation were inclined to argue against the use of vaccines to curb the pandemic and prevent its impacts. This result is concerning but perhaps not surprising in the sense that fringe actors are often more prevalent on social media than in politics in general (Rogers, 2023).

Concerning the sharing of links to news media, all ideological groups rely on partisan media to some extent. This means that selective sharing of news links by choosing sources that confirm and thereby reinforce pre-existing partisan opinions occurs. Here, the Conservative Right sticks out in that it shares links to fringe media websites known for disinformation and far-right views. However, Finland likely differs from countries like the US, where the role played by partisan media is much bigger, as sharing links to the main media sources, such as Helsingin Sanomat and YLE, is common across all groups. However, it must also be stressed that a large proportion of Conservative Right share a different set of sources than the other groups. Plus, our sentiment analysis shows that negative sentiments are over-represented in tweets where links are shared in the Conservative Right group. Thus, the Conservative Right exhibits bias in seeing negative aspects in the topics and interpreting related news in a negative manner. This negativity bias can help to construct a very negative view of, for example, immigration, especially for those whose news diet rests on social media.

More generally, both positive and negative sentiments increase the virality of tweets in which links are shared, but the Conservative Right is again unique; tweets in which negative sentiments are expressed in the context of link sharing get a boost in virality within this group. The Conservative Right thus exhibits a "negativity bias" also in the sense that tweets with negative sentiments tend to become viral (Parkinson, 2020; Schöne et al., 2021). Our results therefore indicate that even though the role of partisan media in Finland is less significant, there are clear partisan differences in how news media links are shared and received on social media.

In interpreting our results, it must not be forgotten that social media users are not a representative sample of the population (McGregor, 2019), as they are likely to be more interested in politics and to hold more extreme opinions than others (Bail et al., 2018). In addition, Twitter (or *X*) is just a single social media platform with its own interaction paradigm and recommendation algorithm. Key areas of future research include both the relationship between the polarisation of Twitter users and citizens in general, and polarisation in (and its potential transfer from) other social media platforms, is therefore needed. Twitter is nevertheless particularly well-suited for studying structural polarisation as a combination of ideological and relational polarisation, over time, and with network methods. Users on this platform also exhibit an interesting mix of politicians, "normal" users, and even conspiracy theorists. By analysing Twitter, we effectively tap into the polarisation of the most polarised segment of society.

Based on our analyses, the cleavage between the Liberal Left and the Conservative Right is noteworthy. In addition, the Conservative Right may be asymmetrically polarised compared to all the other groups in that it exhibits a negativity bias in terms of the news links that are shared and become viral. This attitude may pose challenges for journalism in that conservative-right users are likely to interpret news in a negative light irrespective of what is reported (at least in relation to our main topics). This negative attitude also likely means that the discussion on acute issues, such those that emerge during a crisis, can move beyond rational debate, and impede political compromises. The overly negative attitude, for example, in relation to immigration can potentially be used to create more polarisation by political elites – as did President Trump with his Make America Great Again movement in the United States – or even by a foreign nation bent on creating debate involving strong negative emotions both for and against in the middle of a crisis.

However, we did not analyse the content of the tweets in any detail. In our topics, we are lumping together many "sub-issues" together with the topics and we thus cannot be sure which issues in relation to the topic are most polarising. Future research could thus pay more attention to the evolution of the content of the topic. Further, while we feel confident that the sentiment analysis captures real differences between the groups in how they frame news link sharing, a more detailed content analysis would be useful to see which factors trigger a particular sentiment in a news link sharer.

In addition, future research would be well advised to probe the potential causality between various events reported in the news media and the way in which polarisation on social media reacts to them (e.g., Brodersen et al., 2015). In so doing, it may be possible to develop a framework for explaining what kind of events and news media framings trigger which kinds of responses on social media, and how these responses affect the polarisation of the discussion.

Lastly, a limitation of our study is that in the case of sharing news media links, we did not systematically analyse whether different ideological groups share the exact same versus different links when it comes to sharing links to mainstream media. The systematic differences found in the associated sentiments indicate partisan framings and, in the future, we aim to continue studying the data by analysing whether sentiments differ between the ideological groups *especially* when the groups share the exact same links. One possibility is that differences in sentiments between the ideological groups are especially likely when they share the exact same links. If this is the case, it would mean that sharing the same links to the same media outlets is in fact polarised due to different framings (negative versus positive) and not an indication of agreement between the ideological groups.

APPENDIX

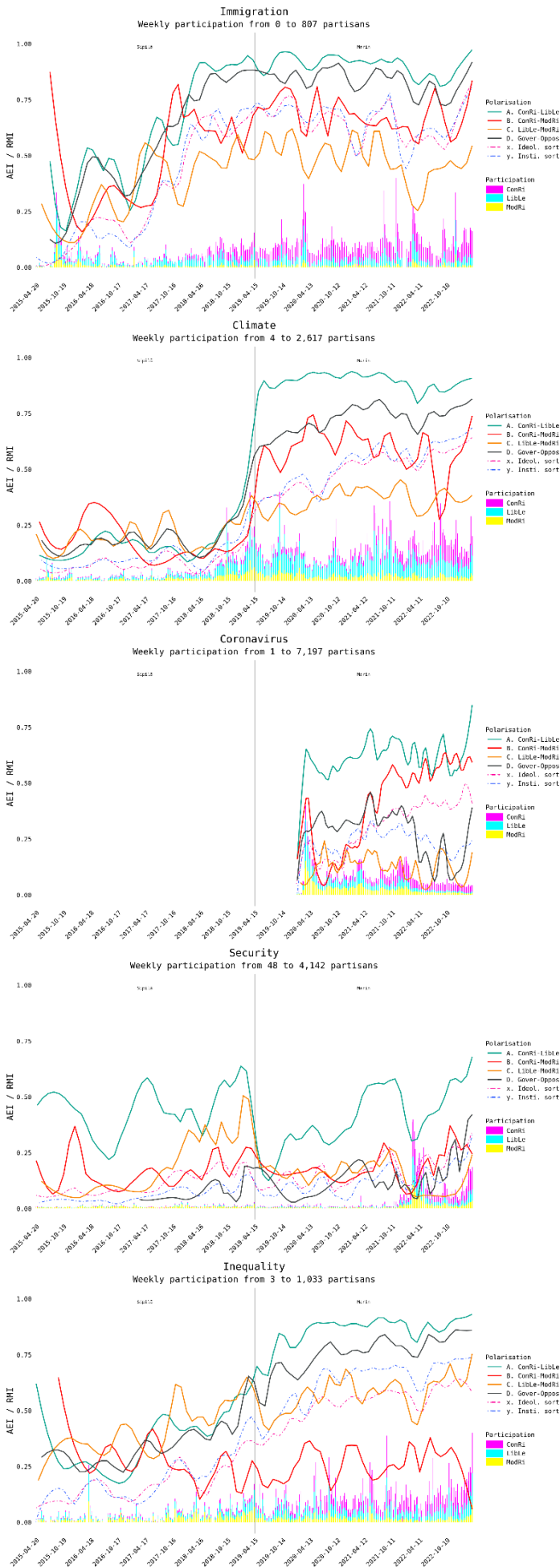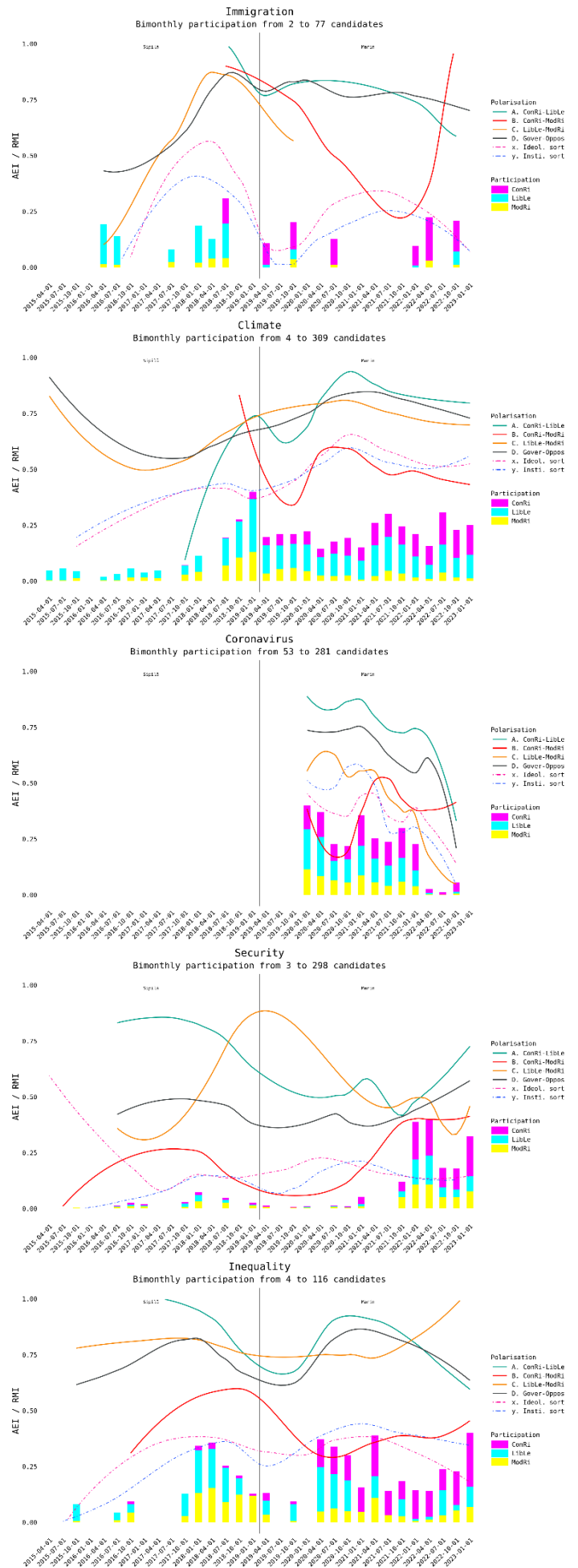

Figure A1. Polarisation for accounts excluding parliamentary candidates, partisan sorting, and participation by topic in weekly intervals from 2015-04 to 2023-03.

Figure A2. Polarisation for parliamentary candidates' accounts, partisan sorting, and participation by topic in bimonthly intervals from 2015-04 to 2023-03.

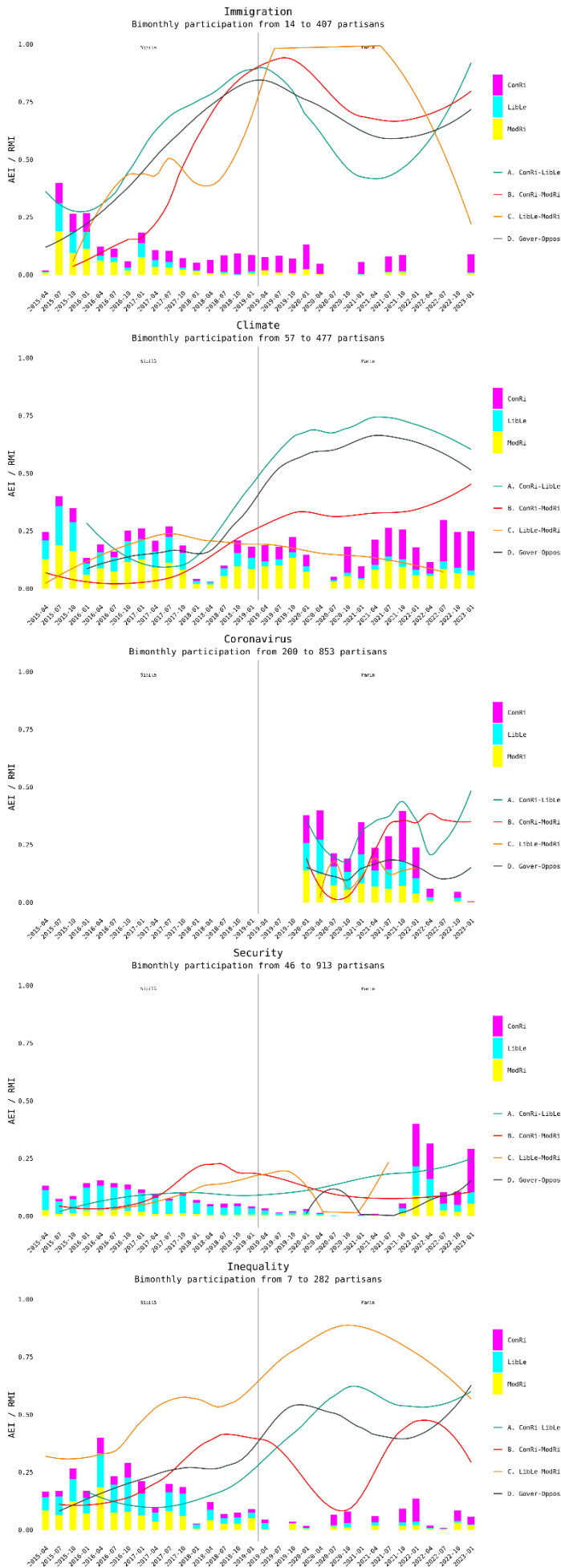

Figure A3. Polarisation for retweets-with-news and participation by topic in bimonthly intervals from 2015-04 to 2023-03.

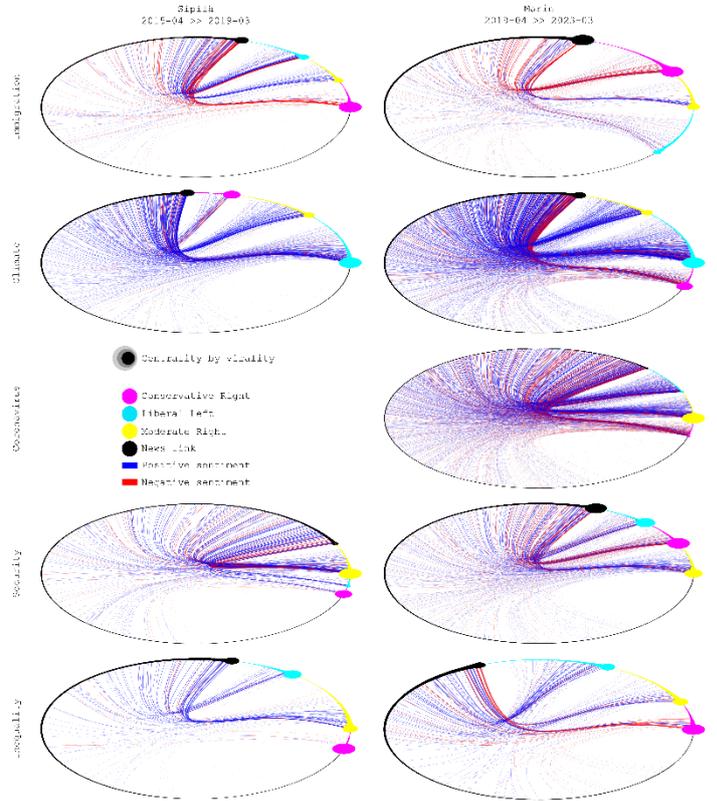

Figure A4. Partisans sharing a news link and reacting to its journalistic content in either positive or negative manner, with neutral reactions hidden from the figure.

*Table A1.* Post-collection substring match keywords by topic.

Immigration
"kansainvälisen suojel", "kansainvälistä suojel", "kansanvälinen suojel", "kansanväliseen suojel", "laiton maahantul", "laittoman maahantul", "maassa maan tavalla", "rajat kiinni", "tilapäinen suojelu", "tilapäistä suojelu", "työperäinen maahanmuut", "työperäisen maahanmuut", "työperäistä maahanmuut", "vapaa liikkuvuus", "vapaaehtoinen paluu", "ankkurilapsi", "haittamaahanmuut", "ihmiskauppa", "integr", "kaksoiskansalai", "kotouttami", "lähtömaa maahanmuut", "maahantul", "maahantunkeutuj", "mamu", "matu", "migri", "muuttoliik", "[*derogatory term for a person of colour*]", "oleskelulu", "oleskeluoikeu", "oleskelustatu", "ongelmalähiö", "pakkopalauttami", "pakolai", "paperiton", "paperittomi", "perheenyhdistämi", "rajatkiinni", "rajaturvallisuu", "refugee", "siirtolai", "säilöönotto", "tphakija", "turvapaikanhakija", "turvapaikanhakijakiintiö", "turvapaikkakiintiö", "kaksoiskansalai", "kiintiöpakolai"

Climate
"extinction rebellion", "vihreä siirtymä", "vihreän siirtymä", "bensakapina", "cleantech", "elokapina", "extinctionrebellion", "fossiili", "geoterm", "greta", "hiilidioksidi", "hiilinielu", "ilmasto", "ipcc", "irtikytkentä", "kasvihuon", "koululak", "lentolak", "lentovero", "lihavero", "lulucf", "metaani", "nytonpakko", "perjantaitgretankanssa", "päästö", "reilusiirtymä", "reilu siirtymä", "reilun siirtym", "thunberg", "tietull", "turbiini", "turpee", "turve", "tuuliturbiini", "tuulivoima", "uniper", "päästöjä", "ydinvoima", "ydinreaktor", "ydinjät", "pienreaktor", "ydinturvallisuu", "atomivoima", "atomireaktor", "atomiturvallisuu", "olkiluo", "ol3", "co2", "fennovoima", "hanhikiv", "aurinkosähkö", "vihreään siirtymä", "aurinkovoima", "tuulienergi", "aurinkoenergi", "ydinenergi"

Coronavirus
"korona", "covid", "pandem", "injektio", "rokote", "rokotte"

Security
"nato", "hävittäj", "hx-hank", "hxhanke", "hornet", "gripen", "f35", "f-35", "krim", "hybridiuhk", "hybridivaikut" "informaatiovaikut", "ottawan", "maamiin", "puolustu", "tp-utva", "tputva", "turvallisuuspolit", "turpo", "palkka-armeija", "sotilasliit", "natsi", "ukropp"

Inequality
"vero", "tuloero", "varallisuusero", "tasa-arvo", "eriarvo", "tuloluok", "tulonsaaj", "minimipalkk", "vähimmäistulo", "vähimmäispalk", "gini", "listaamattom", "rikka", "tulonmuun", "tulojen muun", "pääoma"